\documentclass[12pt]{article}
\usepackage[dvips]{graphicx}

\begin{document}

\title{ Dispersion of nucleation  under the
smooth variation of external conditions}
\author{Victor Kurasov}

\maketitle

\begin{abstract}
A simple method to calculate dispersion
of the total number of droplets
appeared in the process of nucleation
caused by the smooth variation of
external conditions has been presented.
The analytical result for dispersion is
compared with results of numerical
simulations and the coincidence has
been observed. The role of stochastic
appearance of several first droplets in
formation of dispersion has been
analyzed.
\end{abstract}

One of the evidently characteristic
features of a nucleation process is
an occasional appearance of
embryos of a new phase. The
occasional manner of the new
embryos appearance leads to
 generally stochastic corrections in a
 total number of
droplets $N_{tot}$.
In every
particular
process of phase transition the number
$N_{tot}$
differs from the mean value $<N_{tot}>$
averaged over many particular attempts
of nucleation in the given system.
 Since the process
of condensation is essentially non
linear one the mean
value $<N_{tot}>$ which is a result
 of averaging of
particular processes differs from
the value
$ N_{tot\ TAC} $
calculated in the theory   based on
 averaged characteristics
(TAC) \cite{PhysRevE94}.
In a free molecular regime of droplets
growth considered here
the number
$<N_{tot}>$ approaches  to the value
calculated on the base of
averaged characteristics
$N_{tot\ TAC}$. It occurs  when the
volume $V$ of the system
is infinitely big which explains notation
$N_{tot\ TAC} = <N_{tot}(V=\infty)>$.
One has to take this property into account in all
theoretical descriptions of
kinetics of the global process
of phase transition.

Earlier stochastic effects were investigated in
\cite{VestGr}, \cite{Koll}, \cite{Kolldyn},
 but here it
will be shown that the consideration made
in \cite{VestGr},
\cite{Koll}, \cite{Kolldyn} isn't satisfactory.
Still  the structure of analysis proposed in
\cite{VestGr}, \cite{Koll}, \cite{Kolldyn}
can be adopted
in general features
for  further considerations. From the general
limit theorem
for the chain  of
stochastic events it is clear that the
distribution
$P(N_{tot})$
has to be a gaussian distribution
with some mean value $<N_{tot}>$ and
 dispersion $\sqrt{D/2}$.
The task is to determine these values.

From the first point of view $D$ equals
to the value
$$
D_0 = 2 <N_{tot}>
$$
of dispersion of distribution of independent
events.
But the nonlinearity of kinetics of the
nucleation process
leads to the deviation of $D$ from $D_0$.
Namely this
deviation will be the subject of
investigation in this
paper.

Here the situation of the smooth variation
of external
conditions will be studied. Namely in this
situation the
method proposed in \cite{VestGr},
\cite{Koll}, \cite{Kolldyn}
leads to the most striking error.

The plan of investigation will be the
following
\begin{itemize}
\item
In the first part we shall see why the
previous
publications \cite{VestGr}, \cite{Koll},
 \cite{Kolldyn} can
not be applied, what are the errors
made in these
publications.
\item
In the second part a method to determine
dispersion will be
presented.
This method is based on the monodisperse
approximation
proposed in \cite{Various}.
To calculate dispersion one has to
take into account
the specific internal balance property  in
the nucleation
process.
As a result a very simple method to calculate
dispersion is given.
\item
In the third part the
role of the several first droplets in
formation of dispersion is
analyzed. The result of this
section is a simple method
with   clear physical reasons
which can be applied
for practically all significant
relative
deviations of $N$ from $<N>$.
\end{itemize}

\section{Two cycle models of  nucleation process}

The formulation of the problem is described in
\cite{VestGr}, \cite{Koll},
\cite{Kolldyn} in details and
one can omit it here.
To investigate
stochastic effects in kinetics of the
nucleation
process one has to write analytical
approximations to the size spectrum
and then to
variate parameters of these approximations.

The most traditional
way to get analytical approximations
is the iteration procedure.
Originally started by Kuni
\cite{Kuni} it was seriously
modified in \cite{Novos},
\cite{PhysRevE94} and became
a powerful tool to describe nucleation kinetics.
But here one can not use iteration
procedure due to the following reasons:
\begin{itemize}
\item
The simplest
effect which can cause the difference $D$ from
$D_0$ is the influence of
stochastic deviation of the rate
of formation from the mean
 value on the rate of formation
of droplets in next moments of time.
\item
The first
iteration in the iteration procedure corresponds
to the ideal
rate of growth. So, there is no influence of
the deviation of
the rate of formation of droplets of big
sizes on formation of new droplets.
\item
The second iteration can
not be calculated analytically.
\item
The standard approximations
\cite{Kurasov-Vestnik-rect} which
allow to calculate
further iterations correspond to some
model approximation to the size spectrum.
In these approximations one can not
effectively calculate
 the influence of
stochastic deviation of the rate
of formation from the mean
 value on the rate of formation
of droplets in next moments of time.
\end{itemize}

The mentioned
difficulties is the reason to calculate
stochastic effects on the
base of some simple models of
evolution.

To present the models
for the size spectrum one has choose
a renormalization in the
corresponding theory based on
averaged characteristics:
\begin{itemize}
\item
The evolution equation
for the renormalized number of
molecules $g$ in a new phase looks like
$$
g(z) = A \int_{-\infty}^z (z-x)^3 \exp(x - g(x)) dx
$$
Here
$$
\phi \equiv  \exp(x- g(x))
$$
is the size spectrum.
\item
The
characteristic time
$t_*$ necessary for decompositions
(see \cite{PhysRevE94},
\cite{Novos}) is chosen to get at
$z=0$ the maximum if the
spectrum (or the maximum of spectrum
in the first iteration). It leads to
$$
A \equiv 1/6 \approx 0.189
$$
\end{itemize}

The manifestation of stochastic
effects requires to have at
least two effects -
the effect of stochastic deviations
and the effect of
further reaction on these deviations. So,
the simplest model is
the two cycle model where during the
first cycle the
independent stochastic deviations appear
and during the second cycle
the reaction on these deviation
takes place. The
"minimal" model is the two cycle model and
now we shall investigate this model.

At first we shall consider  approximation
proposed in  \cite{Kolldyn}.
The approximation is the following one
\begin{itemize}
\item
Before $z=x_b \equiv 0$
the rate of nucleation is supposed to be the ideal
one.
\item
After $z=x_{b} \equiv 0$
the rate of nucleation $I$ (in renormalized units)
is given by
$$
I  = \exp(x - g_-)
$$
where
$g_-$ is the renormalized number of
molecules in the droplets formed before $x=x_b$
\end{itemize}

In the theory based on the averaged characteristics
$$
g_- = A \int_{-\infty}^0 \exp(x ) (z-x)^3 dx
$$
which explains the sense of approximation.

It is clear that
this approximation is very rough.
The boundary parameter
$x_b$ was chosen in a bad style.
The reasons of such conclusion are the following:

1. The spectrum has very discontinuous character.
The jump at $x_b$ is very
big. It can be seen in fig.1


\begin{figure}[hgh]

\includegraphics[angle=270,totalheight=10cm]{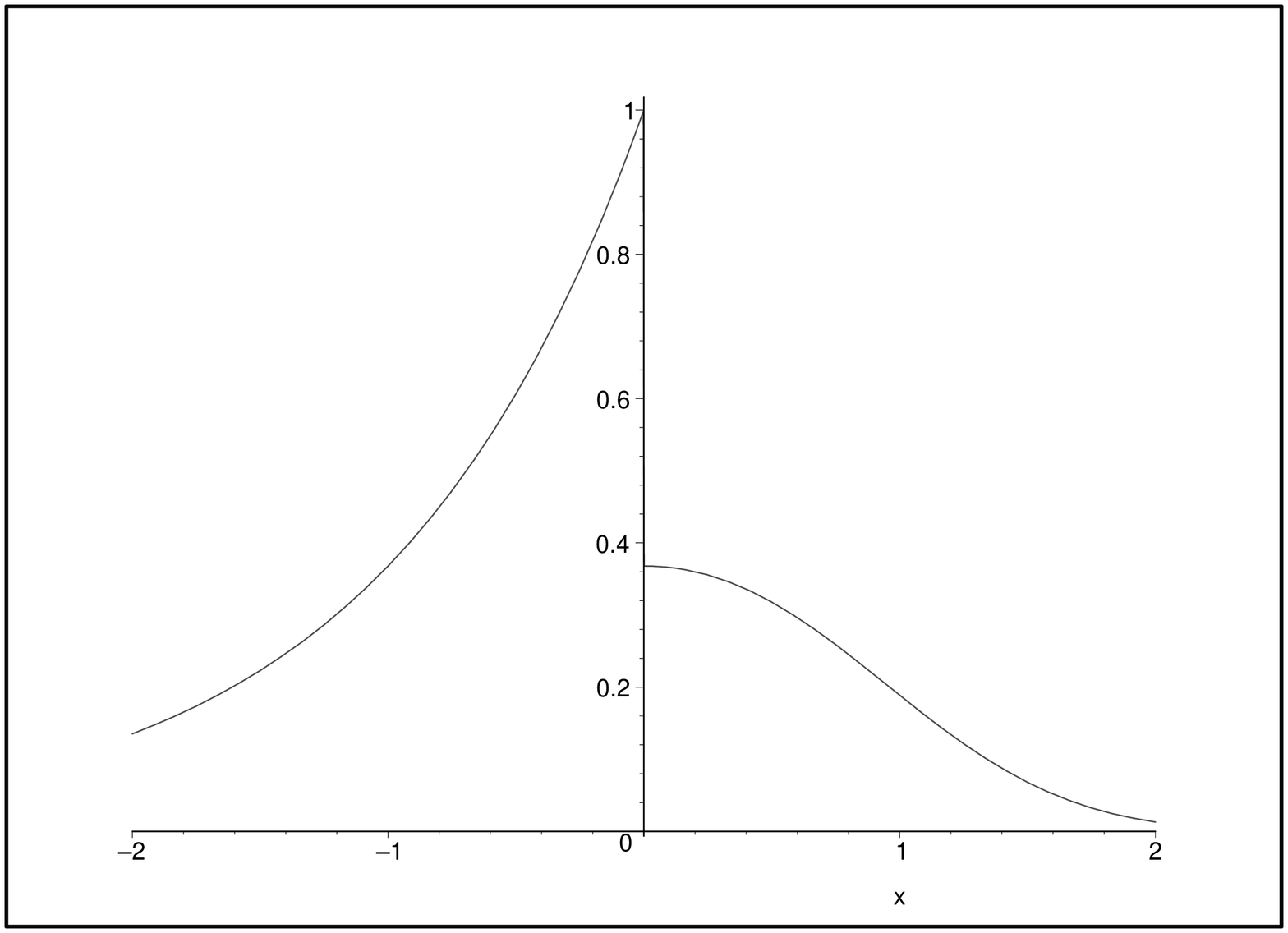}

\begin{caption}
{
The form of  approximation with  a boundary value
$x_{boundary} = 0$
}
\end{caption}

\end{figure}

We see that the region $x>0$ is approximately
negligible in  the calculation of
total number of droplets. But namely
this tail will be the main
source of the deviation of dispersion from the
standard value.

2. For arbitrary $x_{b}$ one can calculate
the number $<N_{tot}(V=\infty)>$
of the total number of droplets.
But always  we get the number
which is greater than the precise value
$<N_{tot}(V=\infty)>  = 1.002$. The
figure 2 shows the number of droplets as function of
$x_{b}$.


\begin{figure}[hgh]

\includegraphics[angle=270,totalheight=10cm]{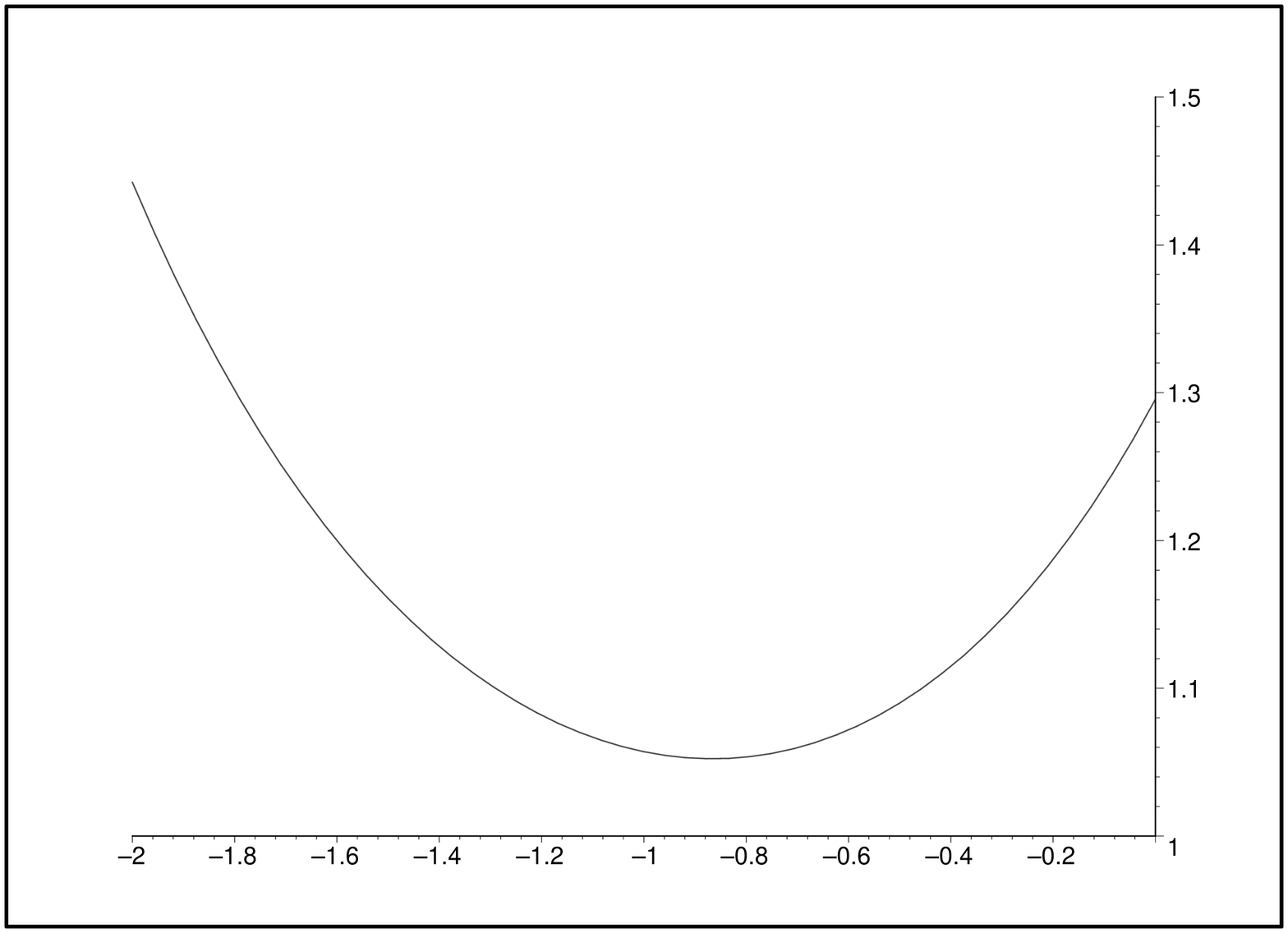}

\begin{caption}
{
The total number of droplets
given by the theory based
on averaged characteristics as a
function of
$x_{b}$
}
\end{caption}

\end{figure}

One can see that the number of
droplets has minimum  at
$x_{boundary} = -0.85$. This value
seems to be more realistic
because the jump in the spectrum
(this jump must be in every
model of such a   type) isn't so giant.
The spectrum is shown in figure 3.


\begin{figure}[hgh]

\includegraphics[angle=270,totalheight=10cm]{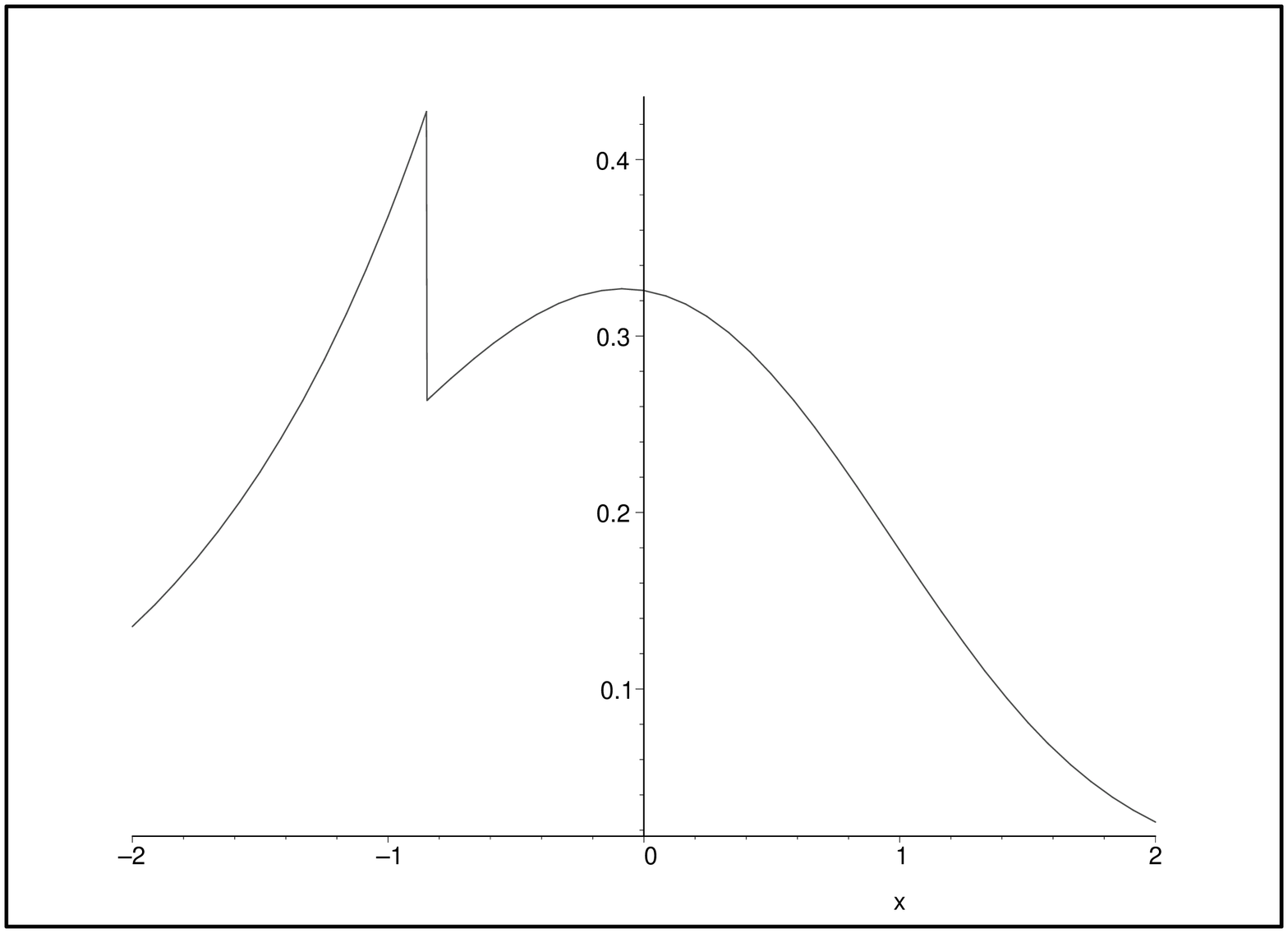}

\begin{caption}
{
The form of spectrum approximation with a
boundary value
$x_{boundary} = -0.85$.
}
\end{caption}

\end{figure}

The models with two cycles of
evolution where the first
cycle is an ideal one and
the second cycle is formed under
the
influence of the first
cycle will be referred to as  "two
cycle models".

The models where the spectrum
in the first cycle is
determined explicitly on the base of
ideal conditions of
the
embryos formation will be called "explicit two
cycle models".

From figure 2 one can see
that explicit two cycle models
can not reproduce the right
number of droplets. But these
models have still more bad
features concerning with
calculation of dispersion.

\section{Numerical calculation of dispersion}

The best way to check the
theoretical calculation of dispersion in the
presented model is to
compare it with the results of
numerical simulation. Here the following type of
numerical simulation will be
presented: The evolution of
the system of the volume $V$
starts from big negative $z$ ($z \approx  - 15$ is
sufficient) and at every step $dz$ the random number
between $0$ and $1$ is
generated $V$ times. Every time the
generated random number is
compared with $f*dz$ where $f$
is the rate of nucleation
in TAC and can be calculated as
$$
f = \exp(z-g(z))
$$
 To know $g$ one has to know four integrals
$I_i, i=0,1,2,3$ and $g$ is determined as
$$
g = I_0 * z^3 - 3 z^2 *I_1 + 3z*I_2 - I_3
$$
The integrals $I_i$ are determined in TAC as
$$
I_i = \int_{-\infty}^z x^i f dx
$$
In application to the random appearance of embryos
  it means that at every
  attempt when the random number
is greater than $f*dz$ the
integrals have to be increased
as
$$
I_i \rightarrow I_i + z^i *1
$$
Certainly, at every step $z$ moves to  $z+dz$.

In the current calculations
the step is $dz=0.001$, at every
$V$ around $1000$ nucleation processes were
performed and then the
mean value and dispersion were calculated.

The interesting value is the
ratio $D/D_0$ which is plotted
in figure 4. Since there is
no evidence what total number of droplets one
has to take
in calculation of $D_0$ (there are two variants
$D_0 = 2 <N_{tot}>$ and
$D_0 = 2 <N_{tot} (V=\infty)>$ )
two curves are presented.
Since $<N_{tot}>$ is very close
to $<N_{tot}(V=\infty)>$
these curves are close. The jumps
of curves correspond to
stochastic nature of calculations
and one can see what characteristic
error is caused by
averaging only over 1000 attempts.


\begin{figure}[hgh]

\includegraphics[angle=270,totalheight=10cm]{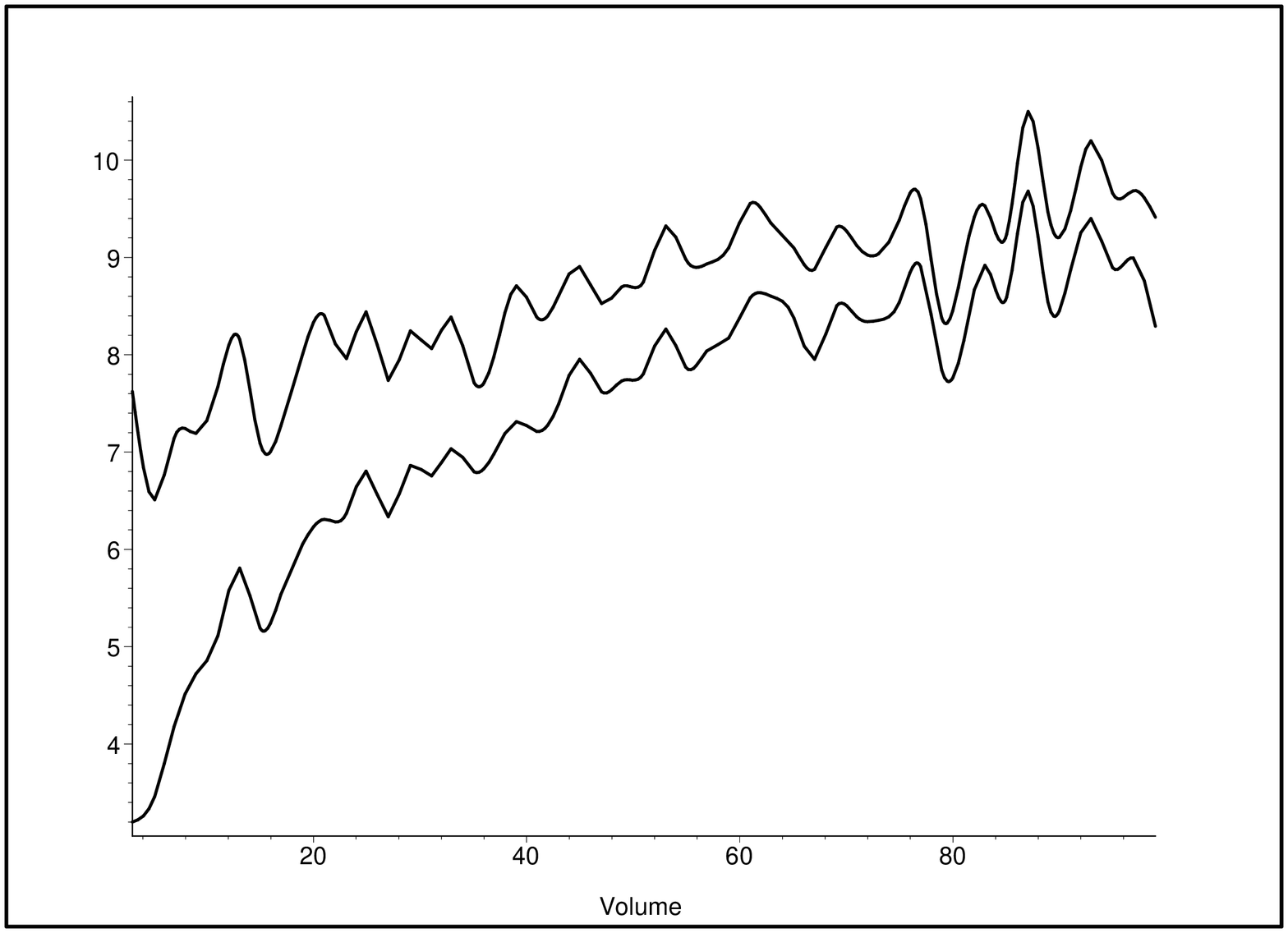}

\begin{caption}
{
Relative square
of dispersion as the function of the volume in precise
simulation.
}
\end{caption}

\end{figure}

One can see that the value of dispersion goes
with increase of $V$ to a limit
value around $9.5$
rather quickly. Really in a volume unit
there will be around
$$
A \int_{-\infty}^{\infty}
\exp(x - g(x)) \approx A = 1/6
$$
droplets.
So, at $10 \div 15$ droplets the limit value is
already attained.

Now one can  turn to the analytical
calculations of dispersion.
At first we shall present
calculation according to
formulas presented in
\cite{Kolldyn}. Since
the approach in \cite{Kolldyn}
is absolutely analogous to
approach \cite{Koll}
we don't mention all mistakes made in
\cite{Kolldyn},
\cite{Koll}. They are mentioned in
\cite{statiaediff}, \cite{statiaediff}
where these mistakes
were corrected.

The result of calculation
is shown in figure 5.
Instead of  \cite{Kolldyn}, \cite{Koll} we consider
the model with an arbitrary $x_b$ instead of
$x_b =0$. Here one can see
the relative square of dispersion as
function of $x_{b}$.

We plot the parameter $\epsilon = D/D_0$
in the Gaussian distribution
$$
P \sim \exp( - \frac{(N_{tot} -
<N_{tot}> )^2}{ 2 \epsilon <N_{tot}> })
$$


\begin{figure}[hgh]

\includegraphics[angle=270,totalheight=10cm]{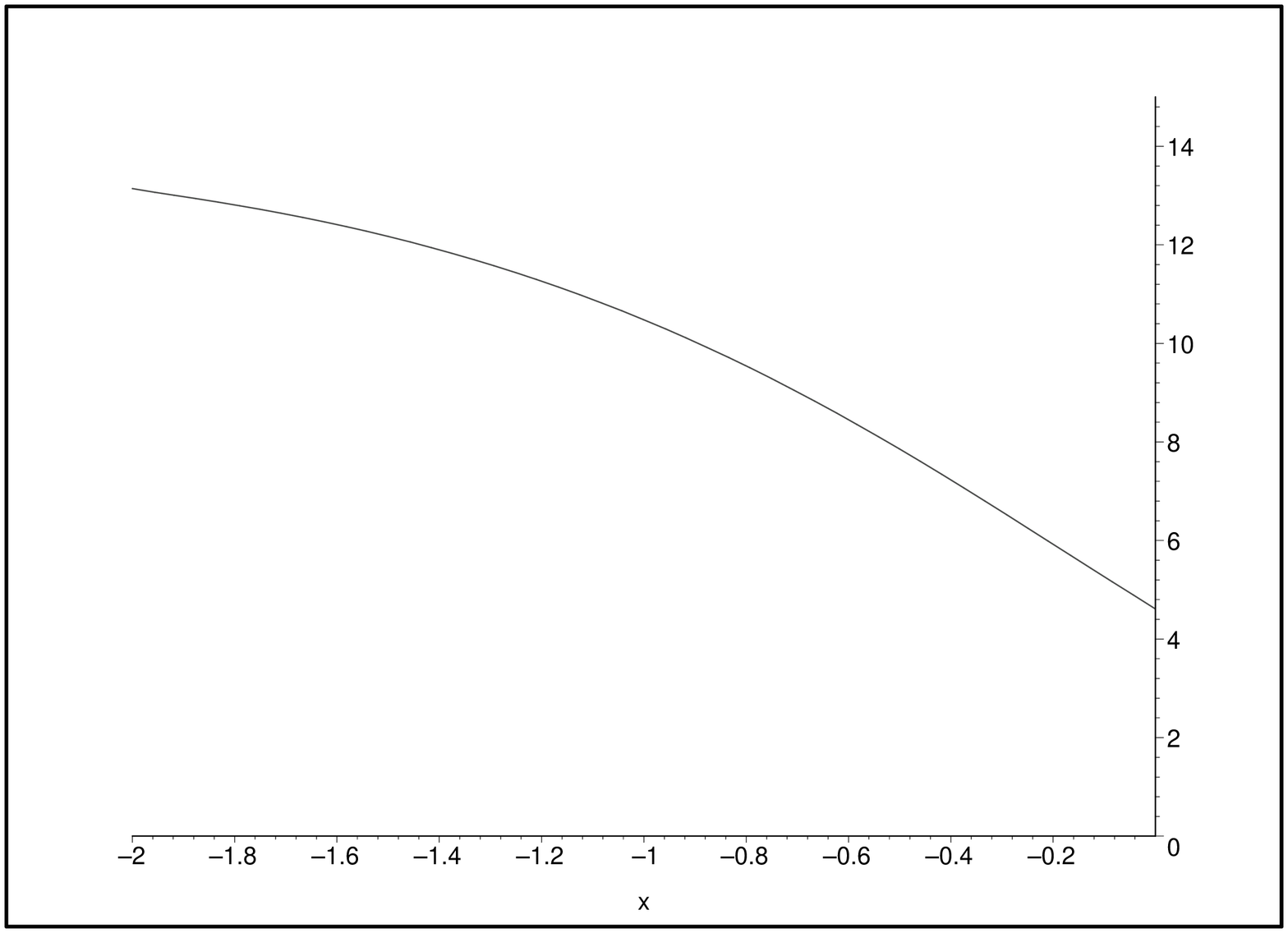}

\begin{caption}
{
Relative square of dispersion as function of
$x_{b}$
}
\end{caption}

\end{figure}

We see that the
dependence of dispersion on $x_{b}$ is
rather essential and
there is no evidence of the choice of
$x_{b}$.

It is seen that the result of
the model with $x_b = 0$ is
two times smaller than a
real result. But the "optimal
choice" $x_b = -0.85$ gives
the value of dispersion which is
rather close to the real result.
It seems that here the
coincidence is attained. But
further considerations will
show that this coincidence in
no more than occasional.

\section{Disadvantages  of two cycle model}

Having seen the results of
two cycle models one has to
mention that  the calculation in
these models is connected
with technical difficulties
and requires serious
simplifications to
come to the final result. Beside many
unnecessary  restrictions (see \cite{statiaediff},
\cite{statiaediff})
the linearization of the action of
stochastic fluctuations of
already appeared droplets on
the
rate of formation of new
droplets has been  done.
This linearization is necessary to
perform analytical
calculations but it
kills all possible deviations of the
mean number of droplets from the
value calculated in the
theory based on averaged
characteristics.  So, the
conclusion about the absence of this shift made in
\cite{Koll}, \cite{Kolldyn} is
out of justification, it is  direct
consequence of linearization.
A special question why
there is no shift in the mean number
of droplets will be the
matter of further analysis and now
the dispersion with all non-linear terms
of the two cycle model will be calculated.

The most simple and effective
way to see the non-linear
corrections in the two cycle model is to fulfill
the
numerical simulation
corresponding to this model. Then
one has to change the
calculation of $f$. Instead of
$f=\exp(x-g(x))$ one has to use
$$
f=\exp(x)
$$ for $x<x_b$ and
$$
f=\exp(x-g_-(x))
$$
for $x<x_b$.
The value of $g_-$ is calculated
analogously to $g$ but the
values of $I_i$ have to be taken
here at $x_b$ (no further
increase is allowed).
So, from the first point of view
there are absolutely no
difficulties in simulation of this model.

The numerical simulation
of explicit two cycle model shows
new weak features of this model.

The first principal difficulty
is the divergence of the
model. Really, there is a
fixed small probability that
until $x=x_b$ no droplets at
all will be formed. This
probability can be taken
from Gaussian distribution as
$$
P(0) \sim \frac{1}{\sqrt{2 \pi <N_{tot}>}}
\exp(- <N_{tot}>/2 )
$$
but here there is no reasons
to take the Gaussian distribution
valid for many
events (here there are no events). It is
more reasonable to
take Poisson distribution but even with
this distribution one has to come to
$$
P(0) \sim  \exp(-l)
$$
Here $l$ is the number of possible events until $x_b$. For
$l$ one can get
$$
l \sim \int_{-\infty}^{x_b}  \exp(x) dx = \exp(x_b)
$$
So, with a fixed
probability (it is very small) there
will be no droplets
until $x_b$. Then $g_- = 0$ and the
size spectrum $f = \exp(x-0)$ grows infinitely.

As a result we see that in explicit
two cycle models there appears a
singularity which is
certainly a main peak in integration
over all possible situations.
So, it completely destroys
the possibility to get
reasonable results in the two cycle
models.

To overcome this difficulty
one has to see what happens
when the number of droplets
$N_{eff}$ appeared before $x_b$
 deviates from the mean value $<N_{eff}>$.
Since the problem is
formulated on the level of the number of
droplets\footnote{It
is really the most important characteristic,
but generally speaking
one can investigate some other
characteristic,
for example $I_i(\infty) = \int_{-\infty}^{\infty}
dx \exp(x-g(x))$.}
it will be very profitable to express
evolution in terms of the
number of droplets. Fortunately
it is possible to do. In \cite{Monodec} a monodisperse
approximation for the evolution in the
theory based on the averaged characteristics
has been presented for the situation of
the metastable
phase decay. In \cite{Various} the
same task was solved for
the smooth variation of external
conditions. Now we shall
use results of \cite{Various} for our purposes.

Briefly speaking one can say that
according to  results of
\cite{Various} during the
nucleation period the vapor is
consumed in the unit volume by $N_0 \equiv
<N_{eff}(V=\infty)>=1/27$ droplets
and this
value in renormalized values is one and the same.

What will happened when at $z=-3$
(this is the average value
for the position of formation of the
maximum of droplets)
the number of appeared droplets
will be less than $1/27$?
The system will wait until
the moment when there will be
$1/27$ droplets in the unit
of volume. It will occur at some
moment later than $z=-3$,
i.e. at $z=-3+\delta$. Later the
evolution will be the same
but in the shifted $z+\delta$.

We see that if we calculate
the number of droplets
precisely at $x=x_b$ then
the parameter in a functional
approximation for the rest of
size spectrum will be smaller
or greater than the necessary
value. In the monodisperse
approximation this functional approximation is
$$
f = \exp(x - N_0 (x-x_b)^3)
$$
In explicit two cycle boundary it is
\begin{equation} \label{pp}
f = \exp(x - g_-(x))
\end{equation}
When we put the smaller $N_0$ than $1/27$ then
the validity of
approximation will be  violated. One simply
can not use
approximation when the value of parameter
seriously differs from the value $1/27$.

The same is valid also for approximation (\ref{pp}).
Approximation (\ref{pp}) can be used
 only if all $I_i$ are
near $\int_{-\infty}^{x_b} x^i \exp(x) dx$.
The  use of
(\ref{pp})
for all deviations (also
 corresponding to the small
values if $I_i$) is the
serious mistake. Namely this
mistake doesn't allow to use the
two cycle models with a
fixed boundaries.

To formulate the two cycle models with
floating boundary
one has to formulate the balancing
property of the kinetics
of nucleation.

\section{Balance property in kinetics of nucleation}

The calculations of stochastic effects
are so complex that they can be made
only in some
primitive models of evolution in frames of mean
values of
characteristics. So, we have to present
simple and rather precise  models of evolution.

At first we consider the situation of
decay.
In the situation of decay there are two sources
of balancing force. The first one is evident.
Suppose that imaginary the whole period is
divided into two arbitrary parts. Let the number
of droplets
in the first part be greater that the average
value. Then these droplets will consume
metastable phase with greater intensity and the
second part of nucleation  will be shorter. Then
the supposed number of droplets born in the
second part will be smaller. This balances the
number of droplets near the average value.

The second balancing reason is more specific.
Suppose that the first droplet isn't going to
appear. What will be changed in kinetics?
Nothing. The start of nucleation process is the
moment of appearance of the first droplet and the
system will simply wait until this moment without
any changes. The process of nucleation is
invariant to the starting point which is
the appearance of the first droplet.

Denote as $t_{st}$ the moment of appearance of
the first droplet. Then
$$
\frac{< d N_{tot}> }{d t_{st}} = 0
$$
in the situation of decay.

Consider now the situation of
smooth variation of external  conditions.
To get the analytic approximations one has to
choose some model of kinetics. The model of
monodisperse spectrum is rather simple and
precise \cite{Various}. We shall follow this model.

Certainly the first reason of balancing
effect takes
place also in dynamic conditions.
 But the second reason doesn't
take place.  The problem is that there is no
fixed initial point such as $t_{st}$ in
decay.

Ordinary the coordinate of monodisperse spectrum
is\footnote{In free molecular regime when the
distribution function
has a maximum at $z=0$.} $z=-3$.
We shall call this coordinate $z_b$.  Then the
distribution function has the form
$$
f = f_* \exp(x- N_{eff} (x+3)^3)
$$
where the effective number of droplets $N_{eff}$
is given
by
$$
N_{eff} = \frac{6 c}{27} , \ \ \ c=0.189
$$
or
$$
N_{eff} = \frac{1}{27}
$$

The value $N_{eff}$ has the gaussian distribution
$$
P(N_{eff}) \sim \exp(-\frac{(N_{eff} -
<N_{eff}>)^2}{2 <N_{eff}>})
$$
where $<N_{eff}>$ is the mean value.

Consider what will happen when at $z=-3$ the
required number of droplets $N_{eff}$ will be
less than\footnote{Certainly, it is
simplification to grasp the main
features.} $<N_{eff}>$. The system will simply
wait until $<N_{eff}>$ will be attained. And then
the evolution will be absolutely the same. But the
evolution will occur
with a shifted value of argument $z$ which
corresponds to the shift in the amplitude value
$$
f_* \rightarrow \exp( k \delta z) f_*
$$
where $\delta z$ is the  shift in $x$  and
$k$ is the coefficient, which can be put
to $1$ by appropriate scale of variables $z,x$.
We keep here $k$ to show that the effects don't
occur due
to the special renormalization.

Then the mean total number of droplets $<N_{tot}> $
will be shifted as
$$
<N_{tot}> \rightarrow <N_{tot}> \exp(k \delta z)
$$

Our task is to obtain the shift $\delta z$ or the
distribution of shifts $\Pi(\delta z)$.
This distribution is connected with $P(N_{eff})$.

One can present $<N_{eff}>$ as
$$
<N_{eff}> = f_* \int_{-\infty}^{z_b} \exp(kx - g(x))
dx
$$
or approximately
$$
<N_{eff}> = f_* \int_{-\infty}^{z_b} \exp(kx)
dx
$$
Then
\begin{equation} \label{tyt}
\frac{d N_{eff}}{dz_b} = f_* \exp(kz_b)
\end{equation}

Then
$$
P(N_{eff}) = \Pi(\delta z)(\frac{d N_{eff}}{dz})^{-1}
= (f_* \exp(kz_b))^{-1}\Pi(\delta z)
$$
or
$$
\Pi(\delta z) = \exp( - k  z_b ) f^{-1}_*
\exp(-\frac{(N_{eff} -
<N_{eff}>)^2}{2 <N_{eff}>})
$$
One can also write the last relation as
$$
\Pi(\delta z) \sim N_{eff}^{-1}
\exp(-\frac{(N_{eff} -
<N_{eff}>)^2}{2 <N_{eff}>})
$$

Then one can speak about the
density distribution $\Xi(N_{tot})$
of the total number of droplets
because
$N_{tot} = <N_{tot}> \exp(k \delta z)$
Then
$$
\Xi(N_{tot}) \sim \Pi(\delta z) k \exp(k \delta
z)
$$

This leads to
$$
\Xi(N_{tot}) \sim P(N_{eff})
$$
and there is no effect of shift. The distribution
of the total number of droplets is
a gaussian distribution with the ordinary mean
value.

The last relation leads to to the absence of shift
 effect
also in the case of dynamic conditions.

Certainly,  one can mention that the absence of
shift
takes place only approximately. Really, having used
in (\ref{tyt}) the same parameter $k$  we
supposed that the difference between ideal
supersaturation and real supersaturation is
absent. But the difference isn't  absent, it
is only small. Then
instead of (\ref{tyt}) we come to
$$
\frac{d N_{eff}} {dz} = f_*
\exp(\frac{\Gamma}{\Phi_*} (\zeta - \Phi_*))
$$
The sense of parameters is explained in
\cite{PhysRevE94}. For us now it is only important
that at $z=z_b = -3$ the coefficient $k$ will be
approximately
$$
k' = k (1-\exp(-3))
$$
Then
$$
\Xi(N_{tot}) \sim P(N_{eff})
\exp( k z_b(1-(1-\exp(-3))) )
$$
or
$$
\Xi(N_{tot}) \sim P(N_{eff})
N_{eff}^{\exp(-3)}
$$

Because $\exp(-3) $ is very small one can  not
see this effect clearly in experiment or in comparison
with numerical simulation.

In fact the existence of
the last effect is rather doubtful and
it
is more preferable to speak about the absence of
 of the
regular shift in the considered situation.

One has to note that the
property of the absence of the regular
shift is very important in the context
of stability of kinetics.
Really all proposed models have to
demonstrate the property of
stability in respect to the
stochastic perturbations. Namely the
absence of the regular shift ensures this
stability.

One can also see that the property
of regular shift will take
place in all two-stage models. The two stage
models of evolution
are the models where the first part of
spectrum is formed under
the ideal supersaturation (it can be the
 monodisperse peak or
explicitly the spectrum like $\exp(x)$)
and the rest part of
spectrum is formed only under the external
 influence and the
influence of droplets from the first
part of the size spectrum.
In details this property will be analyzed separately.

\section{Two cycle models with floating boundary}

The direct result of the
previous balance property is the
absence of correction terms to
the mean number of droplets
formed under the dynamic conditions.
So, the problem to
determine the mean number of droplets
is solved. But
dispersion remains still undetermined.

But now one can formulate the true two
cycle models. They
must be the models with a floating
boundary. The floating
boundary is defined as a
coordinate or the time moment
$z_f$ when the parameters of approximation
concerning the first cycle attain
their values used in
approximation in TAC.
In the monodisperse approximation there will
be no difficulties - parameter $x_f$ in
every particular
attempt is the coordinate
when the number of appeared
droplets attains $1/27$.

Now we shall perform simulations for
explicit two cycle models with
boundaries $x_b =0$ and
$x_b= -0.85$. Results are shown in
figure 6
and figure 7.


\begin{figure}[hgh]

\includegraphics[angle=270,totalheight=10cm]{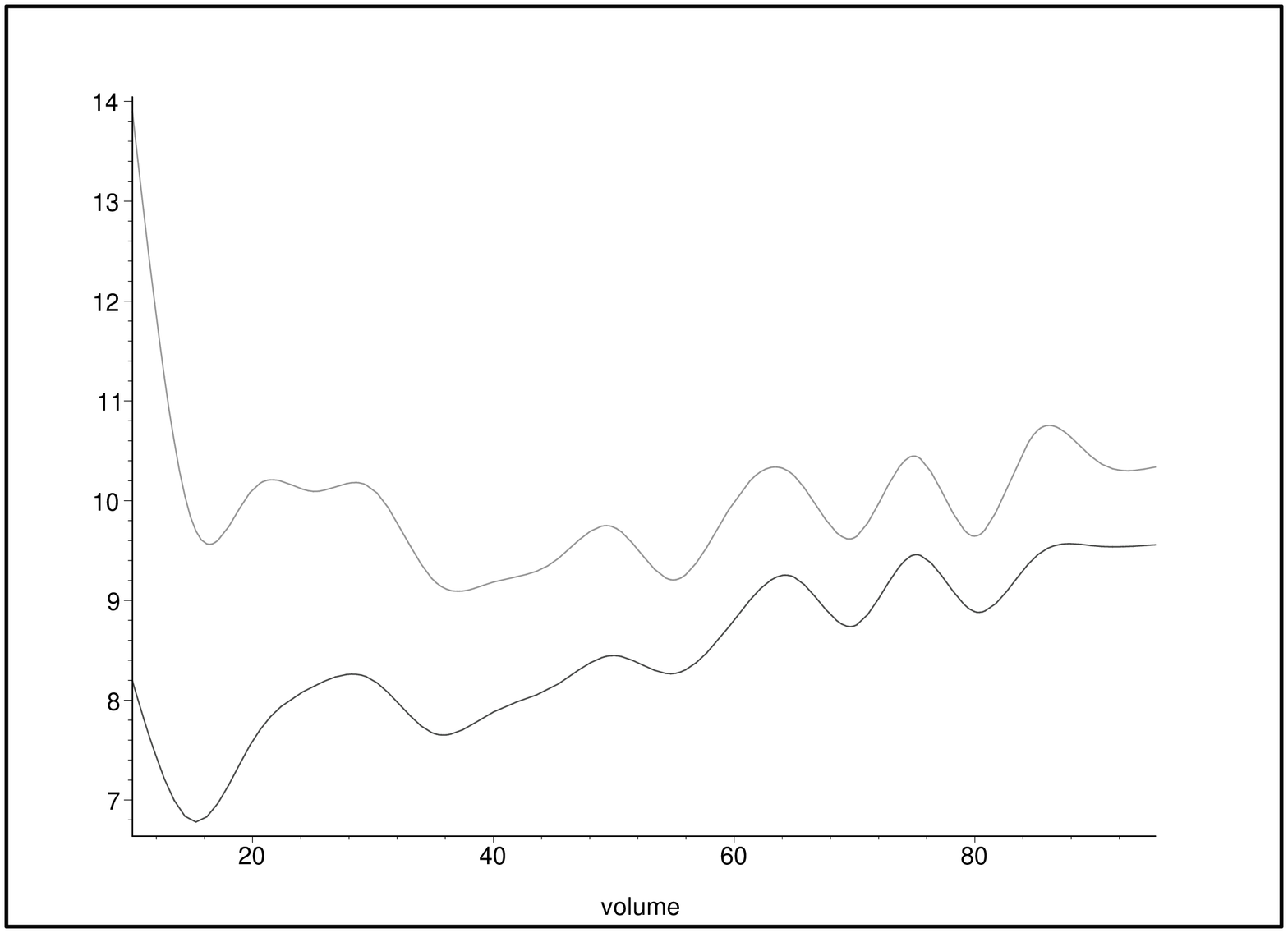}

\begin{caption}
{
Relative square of dispersion as a function of
volume in the explicit two cycle
model with a floating boundary $x_b = 0$.
}
\end{caption}
\end{figure}


\begin{figure}[hgh]

\includegraphics[angle=270,totalheight=10cm]{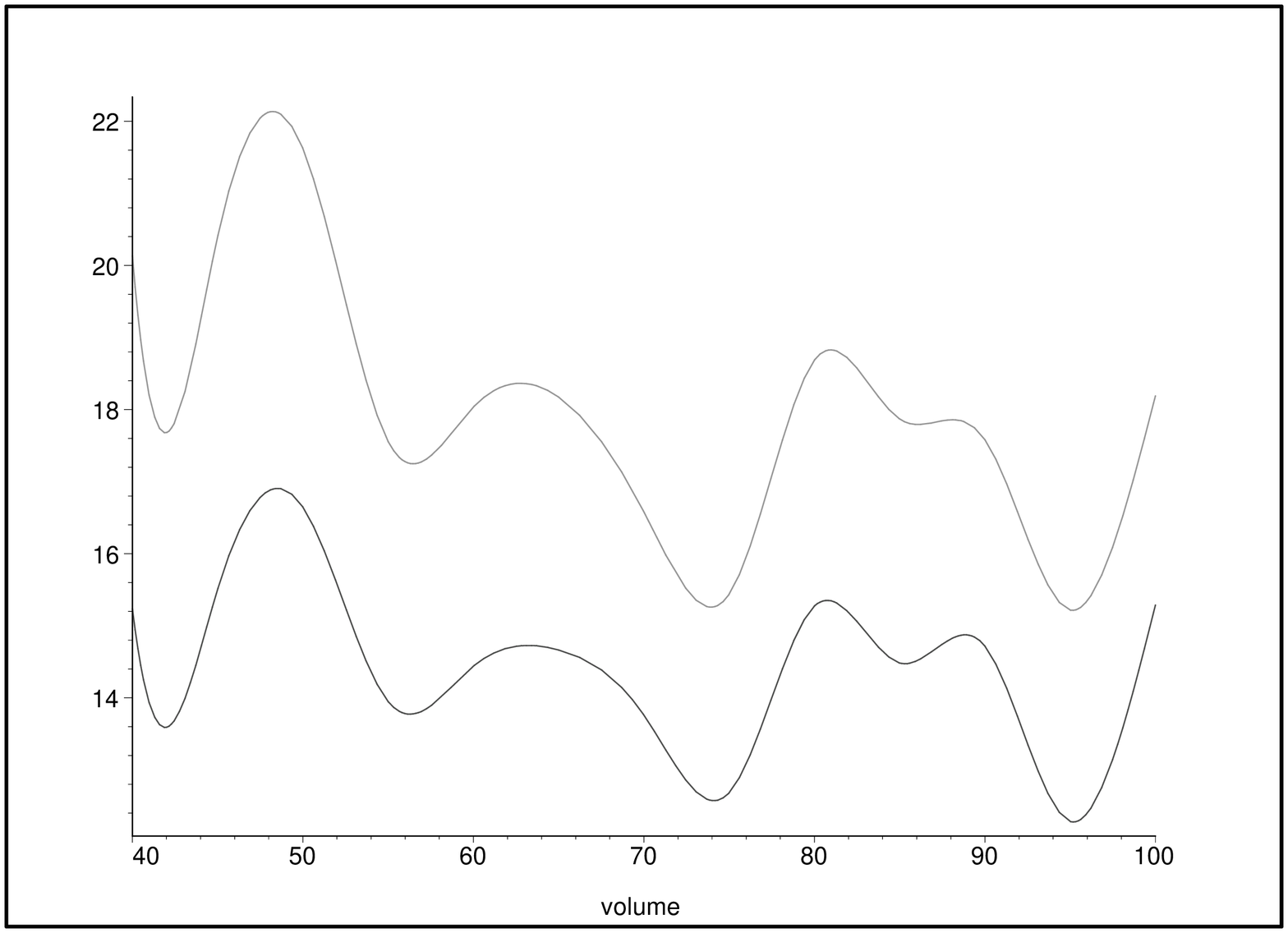}

\begin{caption}
{
Relative square of dispersion as a function of volume in
the explicit two cycle
model with a floating boundary $x_b = -0.85$.
}
\end{caption}
\end{figure}

One can see that the result
of the model with $x_b=0$ is
satisfactory but there are
absolutely no reasons to choose
$x_b =0$ except historical ones \cite{Kolldyn}.
The coincidence is occasional. The most
reasonable model with $x_b=-0.85$
gives a wrong result.

In explicit two cycle models
there will be a problem
because different $I_i, i=0,1,2,3$
can attain necessary
values $\int_{-\infty}^{x_d} x_i
\exp(x), x_{d} = 0;-0.85$
in different moments. This is an
evident  disadvantage of
explicit two cycle models.
Having noticed the weak
features  of explicit two cycle models
mentioned at the beginning, we see
that one has to consider
the monodisperse model with the floating boundary.

The model is the following:
\begin{itemize}
\item
The effective number of
droplets is $N_0 = 6/27$ (the number $6$
appears from $A=6$ in the evolution equation in TAC).
\item
Since the spectrum is formed in an
internal point of a
nucleation period one has
some reasons to suppose that the
subintegral function for $g$
in a current moment of time is
rather symmetrical being related
to the  maximum. Then the
number of droplets which has to be
formed  before maximum has
to be taken as
$N_0 /2$.
\item
The moment $z_f$ of the monodisperse
peak appearance is
characterized by formation
of $N_0 /2$ droplets.
\item
After $z_f$ the spectrum is
$$
f = \exp(x - N_0 (x-x_f)^3)
$$
\end{itemize}

The numerical simulation requires
to specify the procedure
of calculation:
\begin{itemize}
\item
Until $I_0 \geq N_0 / 2 = 3/27$
the function $f$ is
$\exp(x)$ and stochastic appearance of droplets is
initiated.
\item
When for the first time $I_0 \geq N_0 / 2$
the coordinate
$z$ is referred\footnote{When instead of
the mentioned
choice of $z_f$
we choose $z_f$ as coordinate
when $I_0$ attains $N_0$ then
the result of numerical
simulation of this model differs from
result of numerical simulation of the
precise model.} as $z_f$.
\item
In further evolution
$$
f = \exp(x- N_f (x-x_f)^3)
$$
where
$$
N_f = 2*I_0
$$
\end{itemize}

Dispersion in this model is shown in figure 8.


\begin{figure}[hgh]

\includegraphics[angle=270,totalheight=10cm]{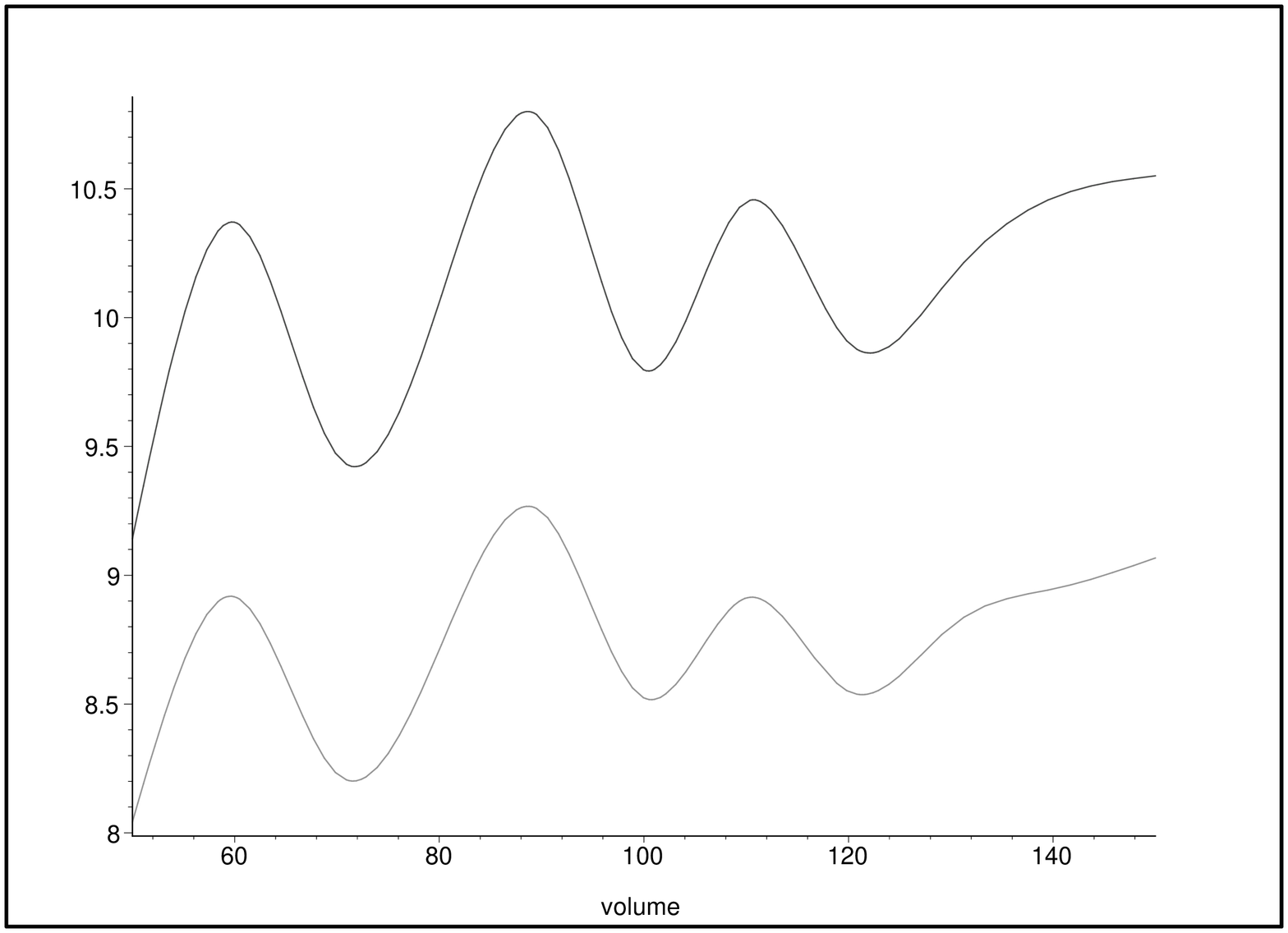}

\begin{caption}
{
Relative square of
dispersion as the function of volume of the
system in the
monodisperse approach.
}
\end{caption}
\end{figure}

One can see that the coincidence
between precise simulation
and simulation in monodisperse
model is satisfactory.

This approximation contains the
precise number of effective
droplets which doesn't
precisely coincide with $1/27$. So, here the
number of effective droplets isn't
the absolutely constant value
prescribed by analytical recipes.
But it is better
to follow this procedure than to
threw out the fractional
part of the number of droplets at
$x_f$. Then even the number of droplets has the
characteristic jumps at the values
of volume $n/27$, where
$n$ is  a natural number.
These jumps a rather essential at
moderate volumes.
It is evident that dispersion
near these values has to be greater
than the ordinary value.

\section{Analytical caluclation of dispersion}

The last model allows very
simple calculations to get the
dispersion of the total number of droplets.
Really, the
dispersion of the first part of droplets, i.e.
the droplets
which are in the monodisperse peak is zero
$$
D_m = 0
$$
since the floating boundary has to be used. The
monodisperse spectrum absolutely governs
further evolution.
It is very easy to calculate the dispersion
of external
influence on further evolution,. Really
according to balance
property and the self similarity of conditions of
nucleation during the nucleation period
(see \cite{Various}) the relative deviation
of the number of rest droplets equals
to  the ideal
dispersion of the distribution of
independent $<N_{eff}>/2$
droplets, i.e.
\begin{equation} \label{ext}
D_{ext} = 2 [<N_{rest}>/(N_{eff}/2)]
* <N_{rest}>
\end{equation}
where
$
N_{rest}$ is the number of remaining droplets
$$
<N_{rest}> = <N_{tot}> - <N_{eff}>
$$
Certainly one has to take into account that under the
fluctuating conditions the droplets are formed
stochastically with internal  fluctuation
characteristic
to the independent
formation of $(<N_{tot}> - <N_{eff}>)$
droplets
$$
D_{int} = 2 <N_{rest}>
$$
Then
$$
D_{tot} = D_{ext} +D_{int}
$$

Having  put $<N_{eff}> = 6/27$,
$<N_{tot}> = 1$ one can
come to
$$
D_{tot} \approx  \epsilon 2 <N_{tot}>
$$
$$\epsilon = 7
$$
The last value isn't too far from
results of simulation.

If one shall take into account the
similarity of conditions
of nucleation
\cite{Various} then it will be
reasonable to put $<N_{rest}> = 1$
(the total number of droplets)
 and then
$$
\epsilon = 10
$$
which is practically the same result as in numerical
simulations.

The advantage of this method is that
it can be used as the
base for further more detailed constructions
taking into
account some fine features of nucleation kinetics.

\section{The role of first droplets in formation of
dispersion}

The effect of the influence of
several first droplets on
the deviation of the mean value
of the droplets number from the
value prescribed by the theory
based on the averaged
characteristics manifests itself
in the fact  that corrections
to the mean value of the number
of droplets can be completely
explained by
formation of several first (actually
the very first one)
droplets. Here such a radical result
does not take place,
but  for a wide range of volumes of
systems one can
effectively use the explicit role  of several
first droplets.

Since several first droplets (at least those in the
monodisperse peak) are formed independently,
the partial
distribution of $n$ droplets is the
Poisson's distribution
$$
P_n  \sim l^n \exp(-l) / n!
$$
Here $l$ is the number
of possible events. Then at the
initial period
$$
l  = \exp(x)
$$
and
$$
P_n (l) dl = P_n (x) dx
$$
It leads to
$$
P_n (x) = \exp( n x) \exp(- \exp(x)) \exp(x)
$$

Certainly, the main
influence appears due to the first
droplet. The partial distribution for the
waiting of the
first droplet is
$$
P_0 = \exp(x-\exp(x))
$$
and it is similar to the universal
distribution observed in \cite{book1}.

The
values of dispersions at
moderate  $V$ initiated by the first droplet,
by two first droplets and by
three first droplets are shown
in the figure 9. The increase of the
number of droplets taken
into account corresponds to the
increase of dispersion -
the
lowest  curve corresponds
to account of only the first
droplet, the highest curve
corresponds to three droplets
taken into account. Here the value of
$\epsilon$ referred to the
number of droplets
prescribed by the theory based on
averaged characteristics is drawn.


\begin{figure}[hgh]

\includegraphics[angle=270,totalheight=10cm]{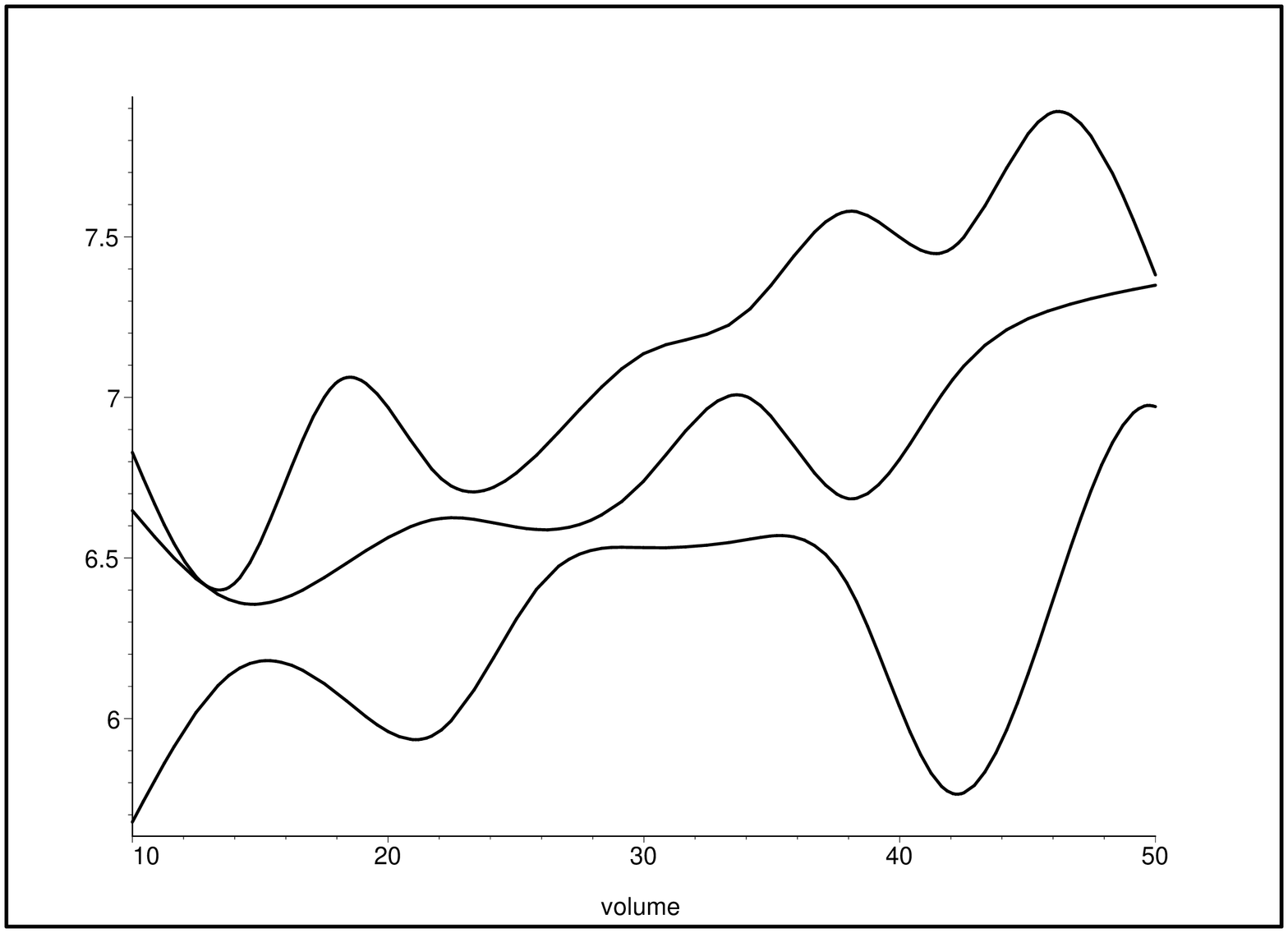}

\begin{caption}
{
Relative
square of
dispersion as the function of volume initiated by
several first droplets.
}
\end{caption}
\end{figure}

It is seen that the increase in  $\epsilon$
initiated by
every new curve is rather small and one can
state that the main
influence is initiated by the first
droplet.

One can see that the
account only of three first droplets
ensures the correct description of the
monodisperse peak
for situations
where $N_{eff} \leq 3$. It corresponds to
the volume $V_3  = 3*27 = 81$
which is already sufficient
for majority of real
practical situation where the
stochastic effects can be
really observed in practice.

\section{Concluding remarks}

The method proposed above is based on monodisperse
approximation. This approximation
is necessary because it
allows to close the system of evolution
equations on a
level of the number of droplets.
The use of explicit models
with more complicated forms
of spectrum used in the
expression for
the quantity of new liquid phase
in corresponding TAC  requires to
write more complicated
equations for other stochastic
characteristics of these
spectrums, which makes the problem
rather more complex.

Meanwhile the applicability of the monodispece
approximation is
based on the sharpness of subintegral
function in
expression for the quantity of new phase. In
the free molecular regime of
growth this function looks like
$(z-x)^3 \exp(x)$ in TAC and
allows the monodisperse
approximation. In the
diffusion regime of growth the
subintegral function in TAC
will be like $(z-x)^{3/2}
\exp(x)$ and can not be
approximated by a monodeisperce
peak so effectively  as
in the case of a free molecular
regime of growth. So,
there appeared some difficulties but
even in this situation the
predictions of monodisperse
model qualitatively coincides
with results of numerical
simulation.

In the diffusion regime of
growth one has to put the
position of monodisperse peak
at $z=-3/2$ and then the
value of necessary droplets in
the monodisperse peak will
be $N_{eff} = 2$ which is
approximately the same as the
total precise number of
droplets in TAC $N_{TAC} = 1.96$.
Then the total number of droplets
appeared in monodisperse
approximation will be $N_{tot\ mono} = 1.98$.
Here $N_{tot \ mono}$
was calculated by the following way:
at $z=-3/2$ the number of droplets
will be the
half of $N_{eff}$
and later the number of droplets is calculated as
$$
N_{tot\ mono} =
N_{eff} /2 +
\int_{3/2}^{\infty} \exp(x -
N_{eff}(x+3/2)^{3/2} ) dx
$$

As the result we see that
$$
N_{TAC} \approx N_{eff} \approx N_{tot\ mono}
$$
It means that
practically all droplets have to be in monodisperse
approximation.
The consequence of this is the absence of
dispersion. But the property that all
droplets are
in monodisperse peak
means that all of them at the same
time govern  their  evolution.
So, then one can
attribute the dispersion
of independent events. Then
$$
D
= D_0
$$
Results of simulation give
$$
D = (1.2 \div 1.25) D_0
$$
which corresponds to above predictions.

Now we
shall refine the model.
Now at $z=-3/2$ the monodisperse
peak is formed and we calculate the number of droplets
explicitly
$$
N_{init} =
\int_{-\infty}^{-3/2} \exp(x) dx = \exp(-3/2)
$$
Then the rest quantity of droplets is
calculated by the
previous formula.

The result of solution
is that approximately the three
quarters of droplets have to be
included in monodisperse
peak and the rest
quarter is formed under the influence of
this peak.
The following
question appears:  shall one takes into
account the dispersion
of droplets in the peak? The
previous result
which shows the self action of the
monodisperse
peak onto itself requires to take into
account the
dispersion of the monodisperse peak.
So we have
$$
D = 3/4 *D_1 + 1/4(1+4/3)*D_0
$$
where
$D_1$ is dispersion
of initial peak. Certainly, for $D_1$
one has to take the
dispersion of independent events $D_0$.
In the last term $1/4(1+4/3)*D_0$ the
value $1$ comes from the self
dispersion of free events and term
$4/3$ is the induced dispersion.

Earlier in investigation of
free-molecular regime there was
no use to take into account the
dispersion of initial peak
because it would give addition only $6/27$.
This addition is so
small in
comparison with renormalizing
factor $10$ appeared
earlier that there is no need
to take it into account. In
diffusion regime such addition  can be essential.

Then the calculation according to the
last formula gives
$$
D = 4/3 *D_0
$$
Then the result is slightly
greater than the result of
simulation but the difference is negligible.

 This result is the
 estimate from above because the
 substitution $D_1 = D_0$ increases the result.
 The first
 estimate $D= D_0$
 is the estimate from below. So,
 $$
 D_0  < D < 1.33 D_0
 $$
 The interval ensured by these estimates is rather
 narrow and the result of simulation lies in
 this interval.

One has to mention that
in the diffusion regime foundations of the
monodisperse
model and of all further
constructions are rather weak. So,
one can not state that
our result isn't well justified.
Fortunately, in the
diffusion regime of growth kinetics of
nucleation is analyzed separately
\cite{PhysicaA} and has
absolutely another physical reasons
of evolution.
Also one has to take into account the
effect of compensation due to the
growing volumes considered in
\cite{statiaediff}.


\begin{thebibliography}{99}

\bibitem{PhysicaA}
Kurasov V.B., Physica A 226 (1996) 117



\bibitem{Novos}
Kuni F.M., Grinin A.P., Kurasov V.B., Heterogeneous
nucleation in
vapor flow, In: Mechanics of unhomogenenous systems,
Ed. by
G.Gadiyak, Novosibirsk, 1985, p. 86
(in Russian)

\bibitem{book1} V.B. Kurasov
Universality in kinetics of the first
order phase transitions, SPb, 1997, 400 p. (in English)


\bibitem{PhysRevE94}
Kurasov V. Phys. Rev. E, vol. 49, p. 3948 (1994)


\bibitem{Koll}
Grinin A.P., F.M.Kuni,
A.V. Karachencev, A.M.Sveshnikov
Kolloidn. journ. (Russia)  vol.62 N 1 (2000), p.
39-46 (in russian)

\bibitem{VestGr}
Grinin A.P., A.V. Karachencev, Ae. A. Iafiasov
Vestnik Sankt-Peterburgskogo universiteta
(Scientific journal of St.Petersburg university)
Series 4, 1998, issue 4 (N 25) p.13-18
(in russian)


\bibitem{Kolldyn}
Grinin A.P., Kuni F.M., Sveshnikov A.M.
Koll. Journ., 2001, volume 63, N6, p.747-754

\bibitem{Kurasov-Vestnik-rect}
Kurasov V., Vestnik SPBGU Ser 4, Issue 3 (N20), p.105-113
(2003)

\bibitem{Kuni}
  Kuni F.M. The kinetics of the condensation under the dynamical
conditions, Kiev, 1984,
-65p. / Preprint Institue for Theoretical Physics Acad. of Sci. UkrSSR:
ITP-84-178E /.


\bibitem{Monodec}

Kurasov V.B., Deponed in VINITI
 Manuscript 2594B95 from  19.09.95, 28p.


\bibitem{Various}
Kurasov V., Preprint cond-mat@xxx.lanl.gov
get  0410043




\bibitem{statiaediff}
Kurasov V., Preprint
cond-mat@xxx.lanl.gov
get 0410616

\end{thebibliography}
\end{document}